\documentclass[aps,prl,twocolumn,showpacs]{revtex4-1}
\usepackage{graphicx}%
\usepackage{epsfig}
\usepackage{amsmath}%
\usepackage{amssymb}%
\usepackage{figsize}
\usepackage{subfigure}

\def\be{\begin{equation}}
\def\ee{\end{equation}}
\def\bea{\begin{eqnarray}}
\def\eea{\end{eqnarray}}
\def\nl{\\ \noindent}
\def\nn{\\ \nonumber}

\begin{document}


\title{\Large\bf How to measure the spatial correlations in disordered Berezinski-Kosterlitz-Thouless transition ?}

\author{Amir Erez and Yigal Meir}
\affiliation{
 Physics Department, Ben-Gurion University of the Negev, Beer Sheva 84105, Israel}
\date{\today}

\begin{abstract}
\noindent The Berezinski-Kosterlitz-Thouless transition is a unique two dimensional phase transition, separating two phases with exponentially and power-law decaying correlations, respectively. In disordered systems, these correlations propagate along  favorable paths, with the transition marking the point where global coherence is lost. Here we propose an experimental method to probe locally these particular paths in superconducting  thin films, which exhibit this transition, and demonstrate theoretically that close to the transition the coherence propagate along a ramified network, reminiscent of a percolation transition. We suggest and calculate experimentally accessible quantities that can shed light on the spatial correlations in the system as it approaches the critical point.
\end{abstract}

\maketitle

According to the Mermin-Wagner theorem \cite{Mermin1966} there can be no long-range order (of continuous symmetry) at any finite temperature in two-dimensional systems. Nevertheless, one may expect a finite-temperature Berezinski-–Kosterlitz--Thouless (BKT) transition \cite{Berezinskii1972,Kosterlitz1973,Kosterlitz1974} between a phase with power-law decaying correlations at  $T<T_{BKT}$ and a phase with exponentially decaying correlations at $T>T_{BKT}$. This BKT transition is driven by the unbinding and proliferation of vortex--anti vortex pairs at the critical temperature $T_{BKT}$, which destroy global phase coherence. It was suggested \cite{Beasley1979} that such a BKT transition can be realized as a  superconductor-insulator transition (SIT) in thin disordered films, where one of the hallmarks of the BKT transition is a universal jump in the superfluid stiffness \cite{Fisher1973} at the critical temperature. This SIT in thin films has been the subject of intense experimental  and theoretical
 research for several decades \cite{Goldman2010}, yet many puzzles remain.  \nl
In such disordered films, it has been established theoretically \cite{Spivak1995,Zhou1998,Galitski2001,Ghosal2001,Dubi2007,ErezMeir2010} and observed experimentally \cite{Kowal1994,Howald2001,Sacepe2008,Sacepe2011} that the superconducting (SC) order parameter fluctuates strongly across the sample, creating "SC islands", where the SC order is high, surrounded by areas of weaker SC correlations. With increasing temperature the coherence between neighboring SC islands is suppressed, until percolation of coherence from one side of the sample to the other is lost, leading to the loss of global SC order \cite{Scalettar1999,Dubi2007,ErezMeir2010}. This description predicts that local SC order may persist even when global SC order is lost, consistent with experiments \cite{Yazdani1995,Gantmakher1998,Crane2007,Stewart2007,Stewart2008,Nguyen2009}. \nl
It has not been established whether the above process is an alternative description of the BKT transition in a disordered system or describes a different phase transition. The Harris criterion \cite{Harris1974} indicates that the BKT transition is stable under weak disorder, and recent experimental \cite{Misra2013} and theoretical \cite{ErezMeir2010} studies on disordered superconductors demonstrated the consistency of the transition in disordered system with BKT physics (see however \cite{Benfatto2007}). The latter calculation also showed that the BKT transition is indeed concomitant with percolation of near-neighbor correlations.

Understanding how coherence propagates in such disordered systems is thus crucial to resolving the nature of the transition. In this Letter, we propose an experimental procedure based on an established experimental technique \cite{Topinka2001,Kiccaronin2004,Jura2007}, whereby the change in a global quantity, such as the resistance or the superfluid response, is measured as a function of a local perturbation, such as a charged AFM or STM tip, which can be scanned across the system. A significant change in the conductance, for example, indicates that the area under the tip carries, in the absence of the tip, a significant portion of the current through the sample, while small changes indicate areas the current tends to avoid. Such a procedure produced clear maps of the modes of transport in quantum point contacts \cite{Topinka2001,Jura2007}  and in the quantum Hall regime \cite{Kiccaronin2004}.  Generalizing the same procedure to SC thin films will allow spatial resolution of  the local propagation of coherence in disordered superconductors, and, more generally, in disordered systems manifesting the BKT transition. (Standard scanning STM measurements have already been done in thin SC films by eg. \cite{Sacepe2008,Sacepe2011}). Here we produce numerical simulations of the experimental procedure on a square lattice, and make specific predictions.
A main result of our work is that indeed certain links carry more of the global coherence as the critical temperature $T_{BKT}$ is approached and that these links form a ramified network. We then present further evidence supporting the percolation hypothesis by showing a connection between this network and the Wolff cluster \cite{Wolff1989}, which is known to percolate at $T_{BKT}$\cite{Chayes1998}. As we demonstrate, the measurement will allow direct probe of local and long-range correlations in the system, which will provide detailed description of the transition. \nl

\textbf{Model and Method:} A minimal model that captures the BKT behavior in two dimensions is the classical XY model. It neglects the fermionic excitations possible when cooper pairs are broken into two quasi-particles, yet it captures the loss of global phase coherence by a BKT transition. This model is relevant to low-dimensional superconductors \cite{Beasley1979}, where loss of phase coherence dictates the transition temperature $T_{BKT}$, which is much lower than the SC gap, the quasi-particle excitation energy \cite{Kivelson1995,Carlson2008}.\nl
The model is described by the Hamiltonian,
\begin{equation}
 \mathcal{H}_{\text{XY}} = -\sum_{\langle i j \rangle}J_{ij} \cos(\theta_i-\theta_j) ,
\label{eq:XY_model}
\end{equation}
With $\theta_i \in [0,2\pi]$ an angle in the XY plane, and $\langle i j \rangle$ denotes nearest neighbors. $J_{ij}$, the random nearest neighbor couplings, are taken from a log-box distribution, $J_{ij}=J_0 \exp(x)$, with $x$ distributed uniformly between $-\lambda$ and $+\lambda$, and  $\lambda\ge0$ thus controls the disorder strength. The distribution is chosen such that the median value of $J_{ij}$ is $J_0$, defined as the unit of energy, regardless of the disorder $\lambda$. In what follows we present results for $\lambda=5$ which provides values of $J_{ij}$ over many orders of magnitude while still allowing enough numerical accuracy, but we find similar results for a range of $\lambda$. We simulate the model using the classical Monte Carlo Wolff algorithm \cite{Wolff1989} with periodic boundary conditions in both directions.\\
The global quantity that we study is the superfluid response, or the helicity modulus. The helicity modulus $\Upsilon_{x}$ in the $x$ direction is the response of the free energy $F$ to a phase twist $\Phi_x$ in that direction and is given by the expression
\begin{eqnarray}
N_x \Upsilon_{x} &\equiv& \frac{\partial^2 F}{\partial\Phi_x^2}
 =\left\langle \sum_{\langle i j\rangle \parallel \vec{e}_x}J_{ij} \cos(\theta_{ij})\right\rangle \nonumber\\
 &-&\beta \left\langle \left( \sum_{\langle i j\rangle \parallel \vec{e}_x}J_{ij}\sin(\theta_{ij}) \right)^2 \right\rangle,
\end{eqnarray}
where the sums are over all bonds in $x$ direction and  $N_x$ is the number of such bonds. $\theta_{ij} = \theta_i - \theta_j$ is the phase difference between neighboring sites $i,j$ and the large brackets denote thermal averages. $\Upsilon_{y}$ is defined similarly for a twist in the $y$ direction, and we define $\Upsilon\equiv(\Upsilon_{x}+\Upsilon_{y})/2$. At low temperatures, where phase fluctuations are suppressed, $\cos(\theta_{ij})\approx 1-\frac{1}{2}\theta_{ij}^2$, and the XY model is equivalent to a random resistor network. Since the resistance of such network with an exponential distribution of resistances is given by the percolating resistance \cite{Ambegaokar1971,Doussal1989}, which, for the square lattice, is the median of the distribution, we expect that $\Upsilon(T)$ will be unaffected by disorder (the value of $\lambda$) for the log-box distributions studied in this work, for low temperatures. Indeed we find that
$\Upsilon(T)$ is unmodified by $\lambda$ at very low temperatures ($T\ll min[J_{ij}]$) and that $T_{BKT}$ is reduced by less than $10\%$ for large disorder ($\lambda=5$). Moreover, the shape of $\Upsilon(T)$ is consistent with the BKT behavior, i.e.  as the system size increases, the (negative) slope in $\Upsilon(T)$ increases for $T>T_{BKT}$ but not for $T<T_{BKT}$, indicating a finite jump in the large size limit.

We simulate a local disturbance (such as the STM tip in the experiment) by cutting a single bond (i.e. setting a single $J_{ij}=0$) and then measuring the resulting change in $\Upsilon$, which we label $\delta \Upsilon_{ij}$. This procedure is repeated for all bonds in the lattice. Removing a bond $\{i,j\}$ in the $x-$direction results in a change in global helicity modulus,
\bea
\label{eq:dY}
\delta \Upsilon_{ij} &=& J_{ij}\left\langle\cos(\theta_{ij})\right\rangle + \beta J_{ij}^2\left\langle\sin^2(\theta_{ij})\right\rangle \nn
&-& 2\beta J_{ij}\sum_{\langle lm\rangle \parallel \vec{e}_x}J_{lm}\left\langle\sin(\theta_{ij})\sin(\theta_{lm})\right\rangle,
\eea
and similarly for a bond in the $y-$direction. As we are interested in linear response, the brackets here denote thermal averages with respect to the unperturbed Hamiltonian in Eq. \ref{eq:XY_model}.\nl

\begin{figure}[]
 \centering
  \subfigure{\includegraphics[clip,width=0.49\hsize]{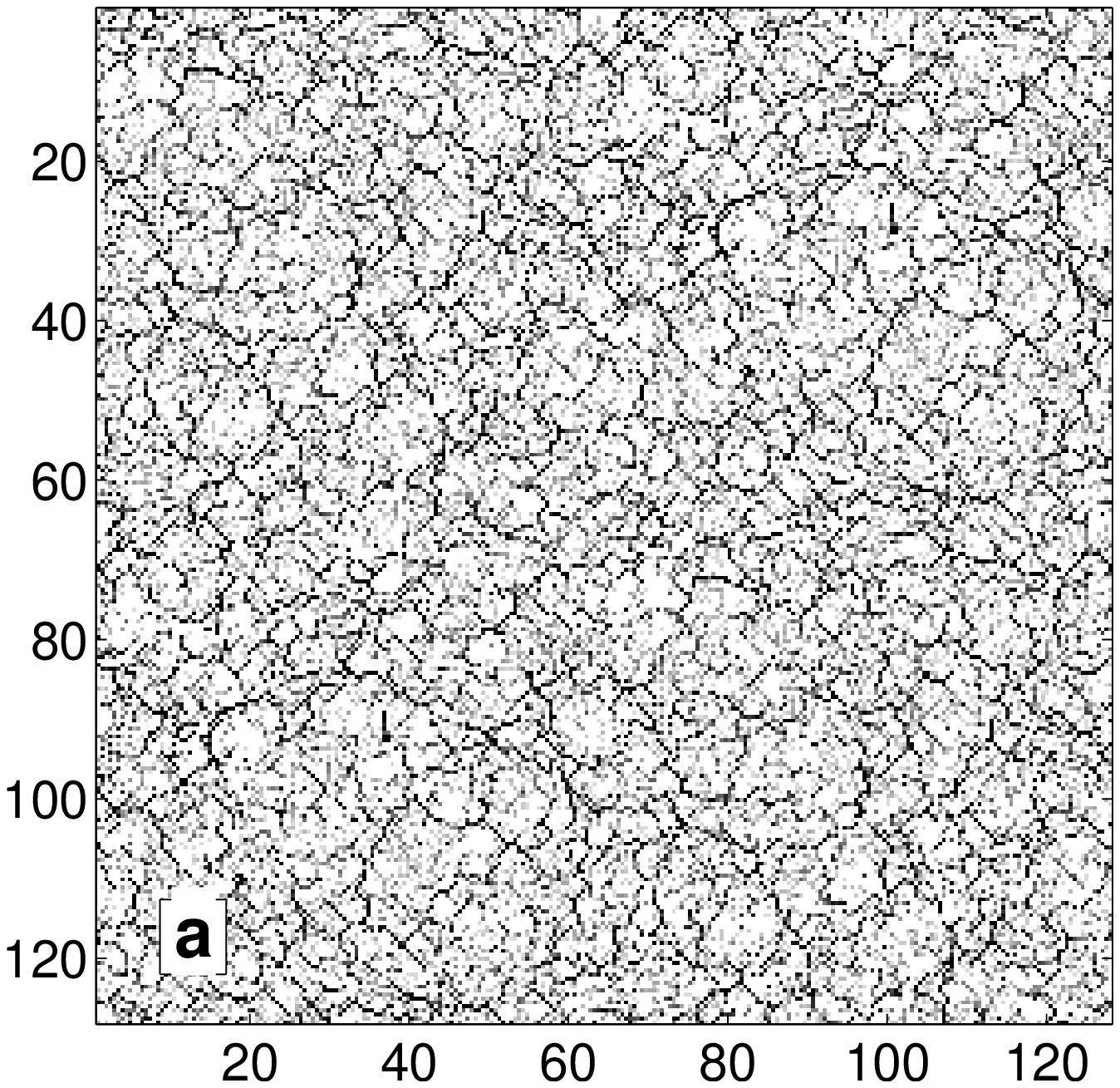}}
  \subfigure{\includegraphics[clip,width=0.49\hsize]{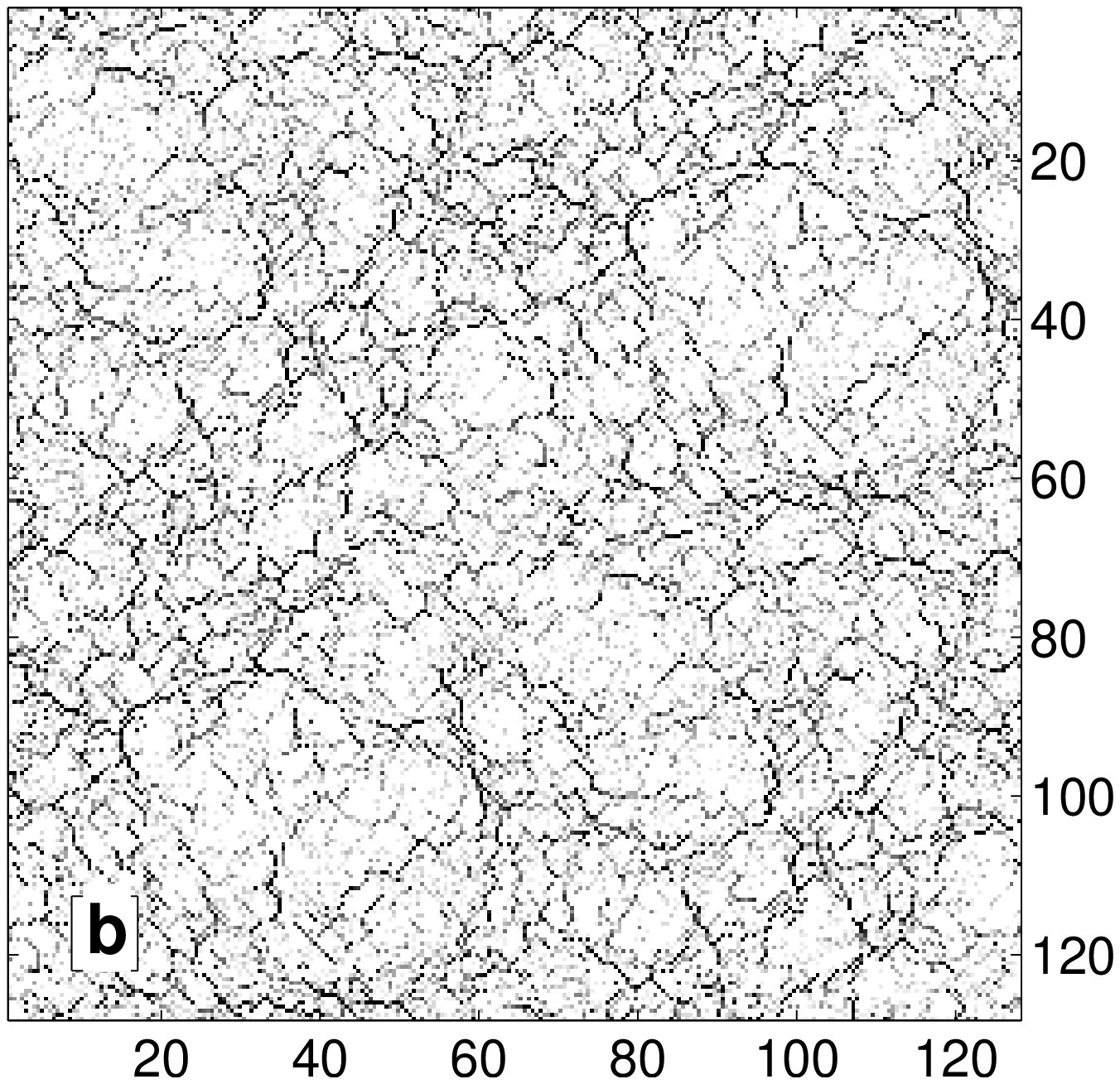}}
  \subfigure{\includegraphics[clip,width=0.49\hsize]{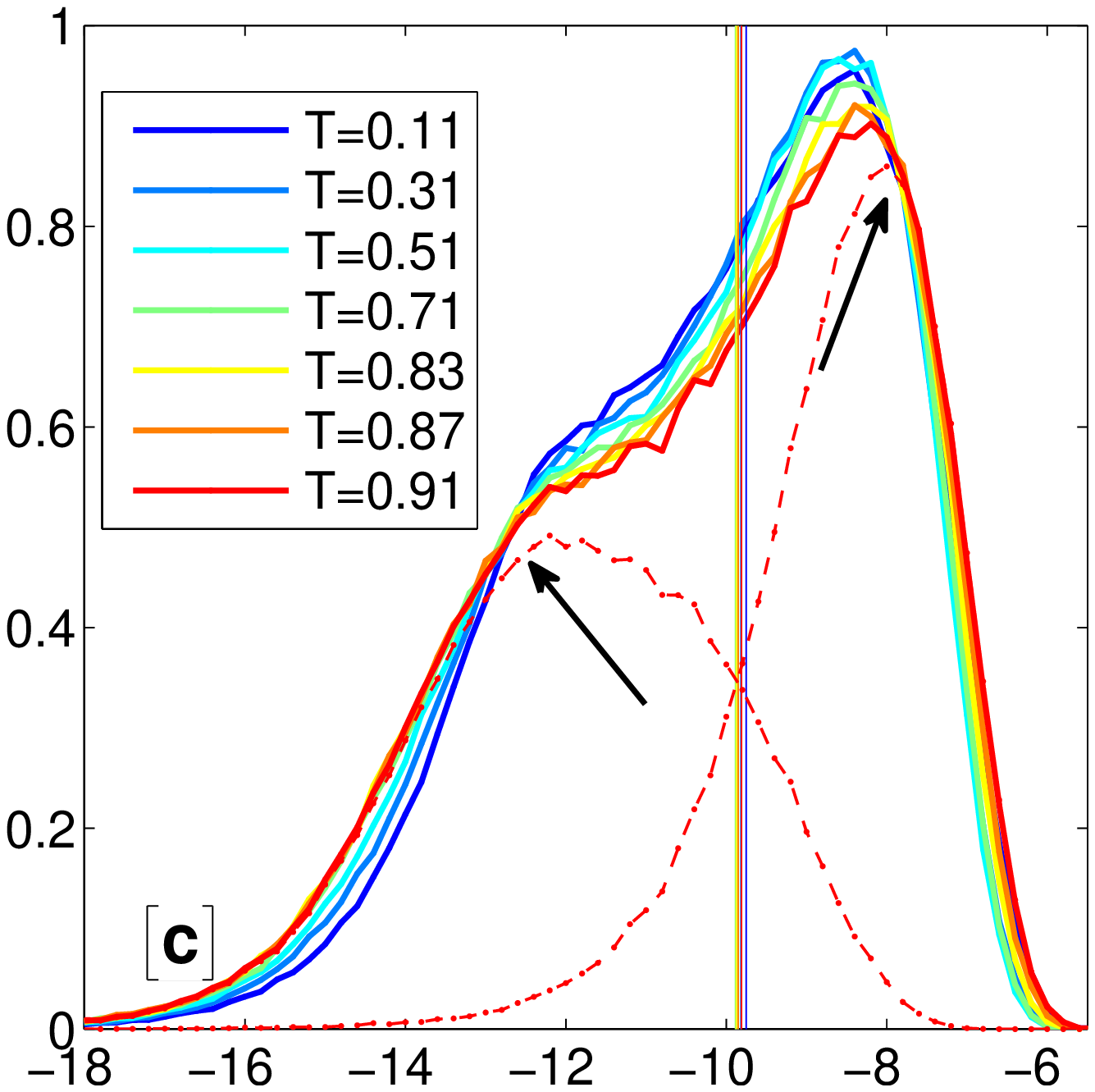}}
  \subfigure{\includegraphics[clip,width=0.49\hsize]{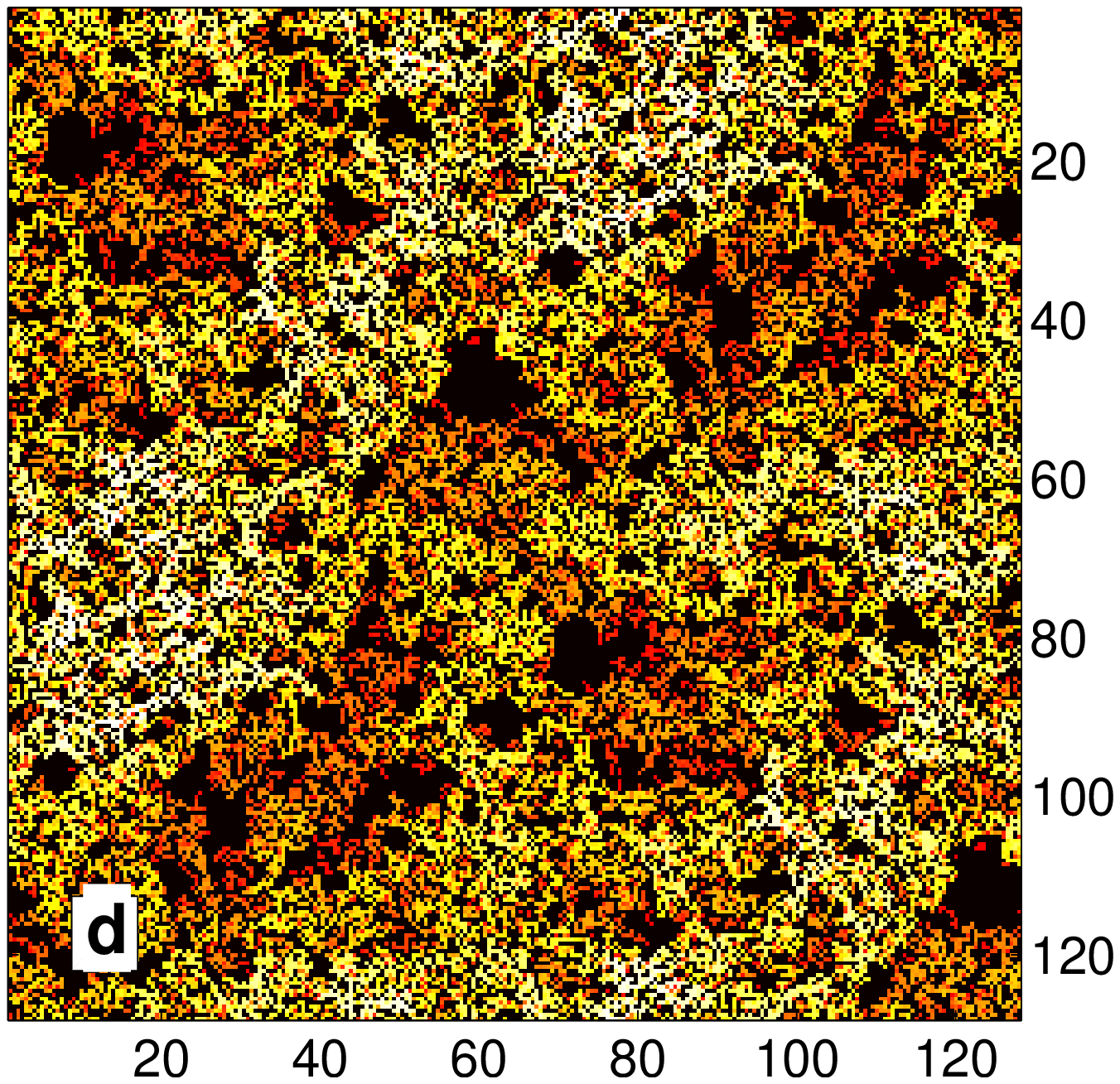}}
  \caption{$\delta \Upsilon$ for a typical realization of disorder at $\lambda=5$ for a $128\times 128$ periodic lattice, for $T=0.11\ll T_{BKT}$ (a) and $T=0.89\approx T_{BKT}$  (b). As temperature increases towards $T_{BKT}$, coherence propagates through specific channels, forming a ramified network. (c) The distribution $P(\delta\Upsilon/\Upsilon)$ at $T=0.11-0.91$ showing a constant median (vertical lines) and a shift of the weight to the tails with increasing temperatures (change points marked by the arrows). Dashed line: the distribution at $T=0.89$ divided to the $J<1$ (left) and $J>1$ (right) components (d) $P_w$ (probability to belong to a Wolff cluster) at $T=0.89$ showing the same spatial structure of the Wolff clusters as the $\delta\Upsilon$ plot in (b).}
  \label{fig:dY}
\end{figure}

\begin{figure}[]
 \centering
  \subfigure{\includegraphics[clip,width=0.49\hsize]{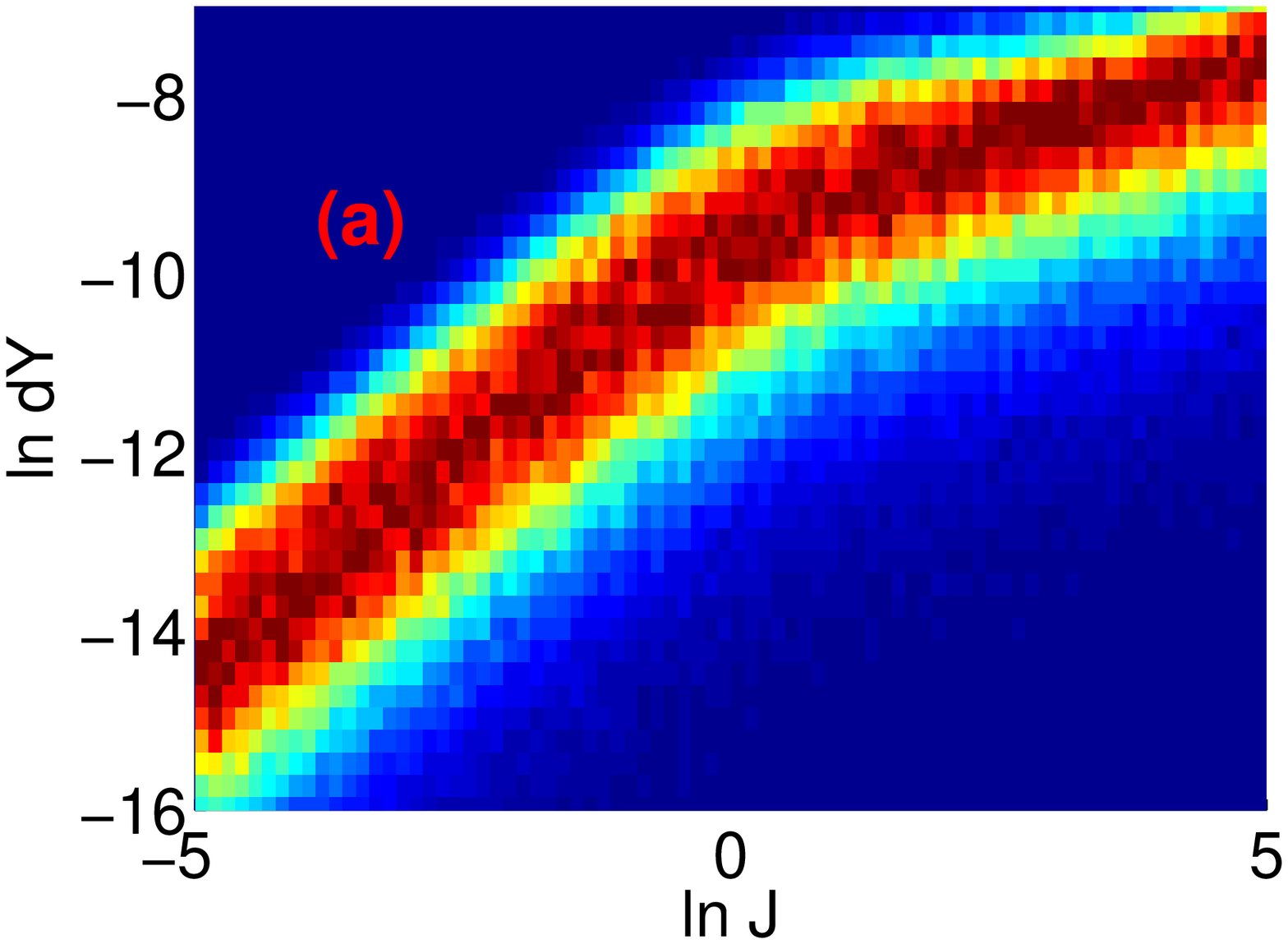}}
  \subfigure{\includegraphics[clip,width=0.49\hsize]{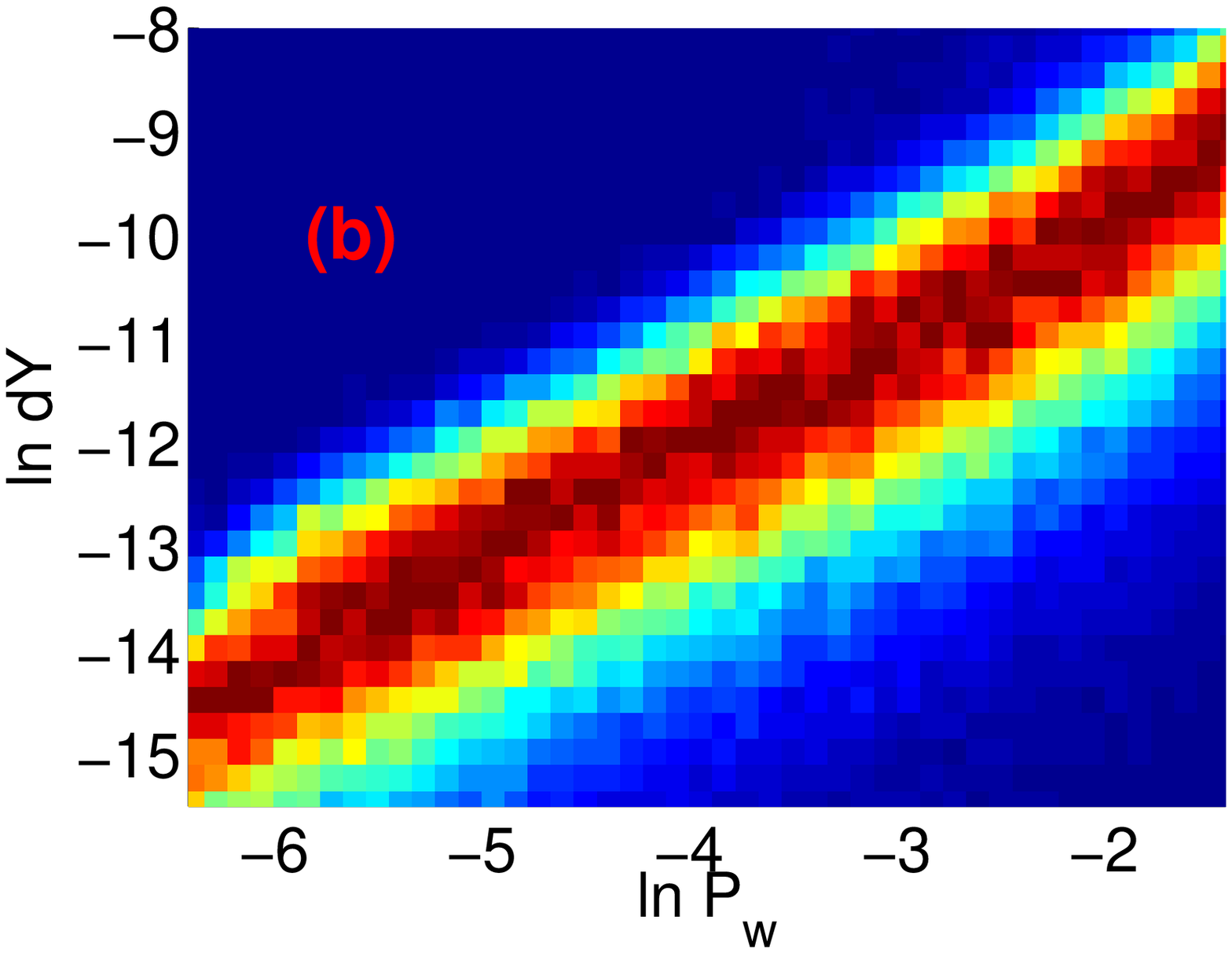}}
  \subfigure{\includegraphics[clip,width=0.49\hsize]{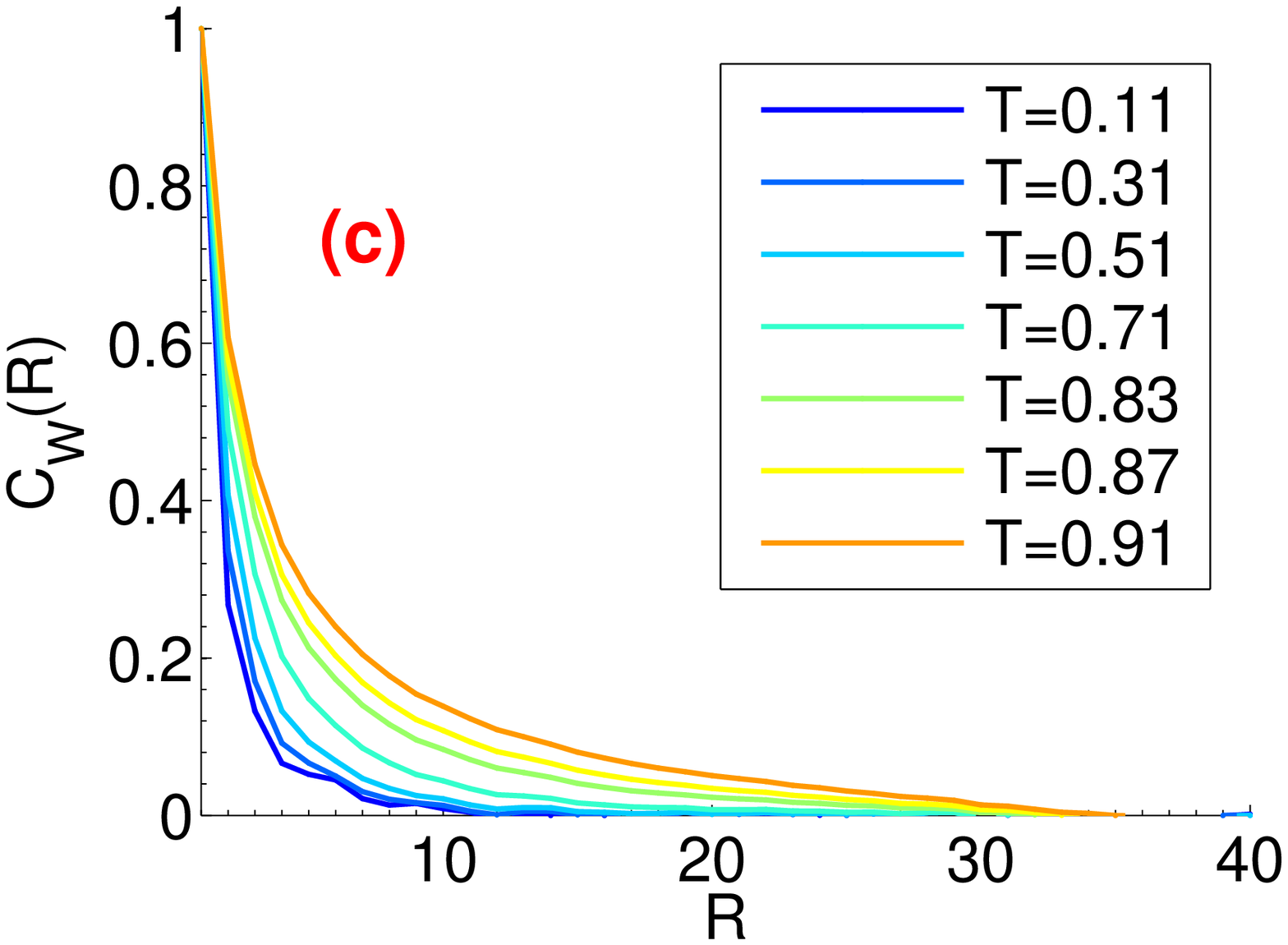}}
  \subfigure{\includegraphics[clip,width=0.49\hsize]{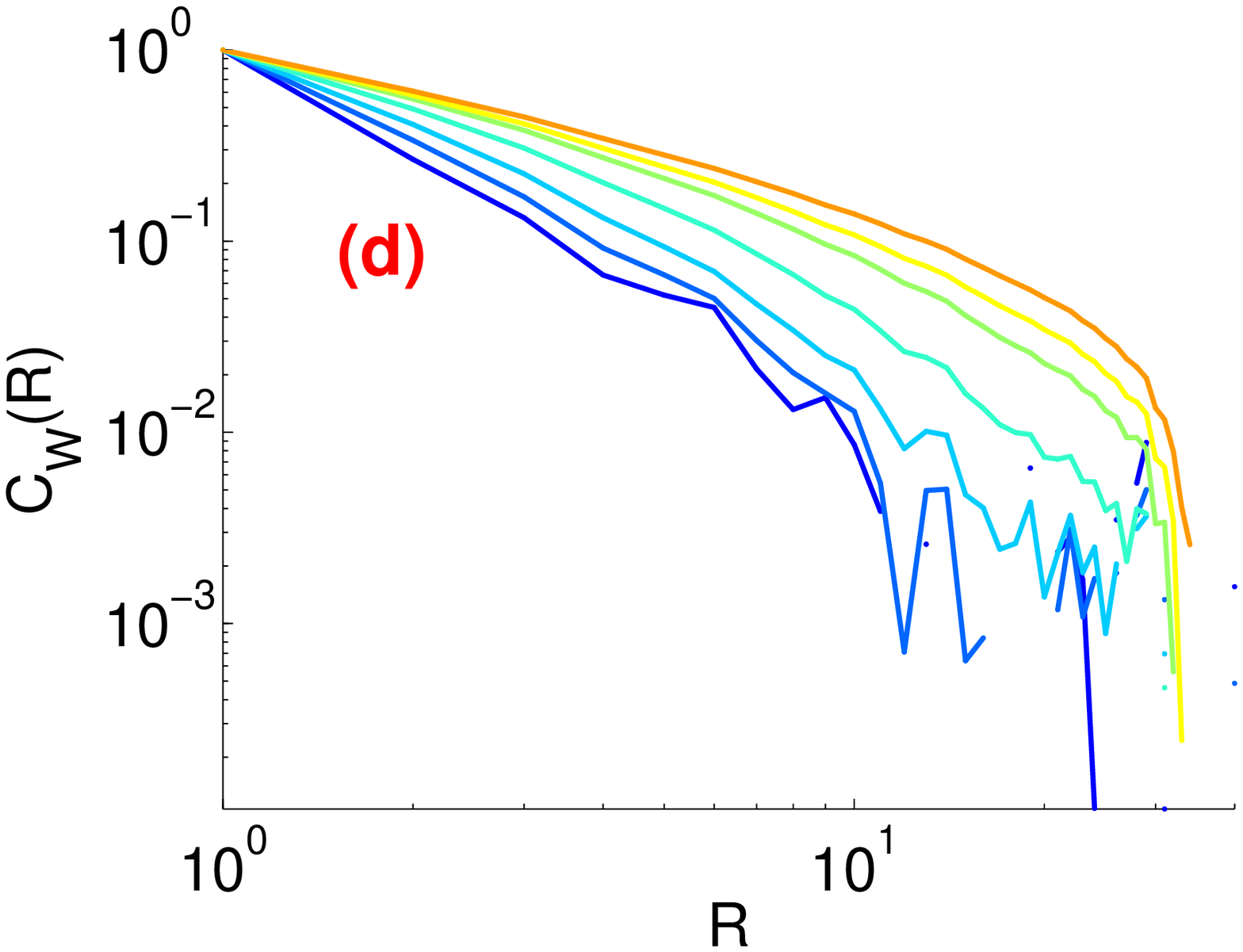}}
  \caption{(a) Normalized scatter plot of $\ln \delta \Upsilon$ vs. $\ln J$ showing correlation only for the sub-median bonds ($\ln J<0$). (b) Normalized scatter plot of $\ln \delta \Upsilon$ vs. $\ln P_w$ (probability to belong to a Wolff cluster) showing a clear linear correlation.(c) $C_W(R)$, the correlation function for $P_w$ for a range of temperatures on a linear and  (d) log-log scale. The correlation decays more slowly at higher temperatures, indicating the sparsity of the coherence propagation. }
  \label{fig:corr}
\end{figure}

\textbf{Results:} Fig. \ref{fig:dY} shows a two-dimensional map of $\delta \Upsilon_{ij}$ for a typical realization of disorder at $\lambda=5$ for a $128\times 128$ periodic lattice, at two temperatures, $T\ll T_{BKT}$ and $T\approx T_{BKT}$. (To display the bond values, we map each bond to a site using the established technique of a covering lattice \cite{Fisher1961}.) Darker bonds mark those bonds that cutting them makes a stronger change to the helicity modulus.
 For $T\ll T_{BKT}$ (Fig. \ref{fig:dY}a) it is evident that the bonds that contribute significantly to global coherence form a dense network.  As $T$ approaches $T_{BKT}$ (Fig. \ref{fig:dY}b),  these bonds form a very ramified network, consisting of distinct channels, even though the disorder does not have any spatial correlations.  In Fig. \ref{fig:dY}c we plot the spatial histogram of the values of $\delta\Upsilon(T)/\Upsilon(T)$ for a range of temperatures.
Three features in the histogram are noteworthy: (i) the median of $\delta\Upsilon/\Upsilon$ does not change with temperature; (ii) As temperature is increased towards $T_{BKT}$, the peak of the distribution (the range of $\delta\Upsilon/\Upsilon$  between the arrows)  gets suppressed and the weight is shifted to the tails of very strong and very weak $\delta\Upsilon$ values. (iii) The distribution is a sum of two peaked distributions, one centered around high $\delta\Upsilon$ (near the right arrow) and one around small values (left arrow). The two broken lines denotes the distributions of  $\delta\Upsilon/\Upsilon$ generated by the bonds with a microscopic $J_{ij}<1$ (left) and those with $J_{ij}>1$ (right), revealing a correlation between $\delta\Upsilon$ and $J_{ij}$ (see below). As disorder ($\lambda$) is increased, the peaks grow farther apart.\nl
Consequently, the following physical picture emerges: at low temperature the sample is mostly coherent and there are many routes for the global phase coherence to propagate through the sample, thus many bonds give a significant contribution when cut. As temperature increases towards $T_{BKT}$, larger regions lose coherence, so the coherent routes gradually thin out, limiting the propagation of global coherence to a smaller set of bonds, forming a ramified network of channels.\nl
Naively, one might suspect that there exists a trivial relation between the bare microscopic couplings $J_{ij}$ and their global effect $\delta\Upsilon_{ij}$, as might be inferred from Eq. \ref{eq:dY}. Yet Fig. \ref{fig:corr}a reveals a more interesting relation: for sub-median bonds $J_{ij}<1$, there is a linear correlation $\delta\Upsilon_{ij}\propto J_{ij}$ as suggested by the unit slope of the log plot. However, for half the bonds of the lattice, for which $J_{ij}>1$, $\delta\Upsilon_{ij}$ is only weakly dependent on $J_{ij}$ (the exact relation depends on $\lambda$). The reason is that these bonds belong to the percolating coherence cluster, and the effect of cutting such a bond on $\delta\Upsilon_{ij}$ depends no longer only on its bare coupling $J_{ij}$ but rather its location in the network. We consider this behavior and its onset at the median bond (which is the percolation threshold for the square lattice) as further evidence supporting the percolation scenario.\nl
In order to shed more light upon the spatial structure of the correlation, we turn our attention to the Wolff clusters \cite{Wolff1989}. A priori, these are  mere numerical constructs that were introduced to overcome the numerical problem within the Metropolis Monte Carlo algorithm, which is based on updates of local variables (such as spins), depending on Boltzman weights. Due to the divergence of length scales near the transition, spins are highly correlated on large scales, and single spin flips are highly improbable. To address this problem Wolff \cite{Wolff1989} proposed a very efficient algorithm which eliminated this critical slowing down by allowing clusters of spins, called the Wolff clusters, to flip together. These clusters
are constructed stochastically according to near-neighbor correlations.
It was later suggested \cite{Henley1991} and then proven rigorously \cite{Chayes1998} that the efficiency of the algorithm lies in its correct capture of the physics of the BKT transition. The size of the Wolff clusters, that reflect the correlated spins, grows towards the BKT transition, the latter happening exactly where the Wolff cluster percolates through the whole system. Note that this proof applies for a {\sl clean} system - the random structure of the Wolff cluster is a result of the stochastic nature of its construction, so one can imagine that they reflect the random temporal correlations of the spins. Nevertheless, due to the uniformity of the system, the probability of every bond $<ij>$ to belong to a Wolff cluster, $P^W_{ij}$  is the same. This situation is very different in disordered systems. The Wolff clusters, generated during the Monte Carlo calculation, are highly anisotropic and reflect the spatial disorder in the system. In Fig. \ref{fig:dY}(d) we display a two dimensional plot of  the probability $P_{ij}^w$ at $T \approx T_{BKT}$. Indeed, $P_{ij}^w$ and $\delta\Upsilon$ (Fig. \ref{fig:dY}(b)) exhibit very similar spatial structure, so the bonds with high $\delta\Upsilon$, darkest in Fig. \ref{fig:dY}(b), are also the ones with the highest probability to be part of a Wolff cluster, brightest in Fig. \ref{fig:dY}(d). Hence the bonds that support the propagation of coherence  (i.e have a large $\delta\Upsilon$) are also, as expected, the ones that have long range correlations, as reflected by the Wolff clusters.  This observation is further supported by Fig.\ref{fig:corr}(b), which displays a normalized scatter plot of $\delta\Upsilon_{ij}$ vs $P_{ij}^w$, demonstrating the clear correlation between over ten orders of magnitude. If the proof that the Wolff clusters percolate at the BKT transition extends to the disordered case, as is indicated numerically, it is yet another support for percolating nature of coherence at the BKT transition.\nl
A further check on the changing spatial structure with temperature can be revealed by studying the correlation function
\begin{equation}
C_{\Upsilon} (R)\equiv \frac{\sum_{|<ij>-<kl>|=R} \left(\delta\Upsilon_{ij}-\overline{\delta\Upsilon}\right)\ \left(\delta\Upsilon_{kl}-\overline{\delta\Upsilon}\right)}{\sum_{|<ij>-<kl>|=R}} ,
\label{eq:CY}
\end{equation}
where the sum  is over all pairs of bonds, distance $R$ apart, and $\overline{\delta\Upsilon}$ is the asymptotic average of $\delta\Upsilon_{ij}$, so that $C_Y (R)$ decays to zero at large $R$. A similar correlation function $C_W (R)$ can be defined in terms of $P_{ij}^w$. Due to the high correlations between $P_{ij}^w$ and $\delta\Upsilon_{ij}$ (Fig.\ref{fig:corr}(b)) we find a similar behavior  for these two correlation functions, though  $C_W (R)$ displays smaller statistical fluctuations. Fig.\ref{fig:corr}(c) depicts $C_W (R)$ at a range of  temperatures and Fig.\ref{fig:corr}(d) depicts it on a log-log scale. At low temperatures, as the network supporting coherence is dense, $C_W(R)$ decays quickly to zero. As temperature increases there is a marked slower decay of  $C_W (R)$, demonstrating the ramified nature of that network. For the system sizes we simulated, the correlation function doesn't follow any simple decay law, possibly due to finite size effects. 

\textbf{Summary and discussion:} The effect of disorder on the BKT transition is a hotly debated problem in condensed matter physics, and its relation to percolation, or the emergence of a new critical point are open issues.  In this work we propose an experiment to probe the spatial structure of global phase coherence in two dimensional superconductors, that are expected to display such a transition. Our proposition is based on similar experiments that probe conductance channels through quantum point contacts and in quantum hall systems (while, for theoretical convenience, we display results for the phase stiffness, we expect similar results for the current). We present numerical results for a minimal model that (in its clean version) displays a BKT transition, in the strong disorder regime. Our numerics point at an intriguing mechanism where the global coherence passes predominantly through specific channels that can be experimentally probed. The correlation functions we have calculated, which can also be measured experimentally, also support the idea that as the critical temperature is approached, the coherence network becomes more ramified, with percolation lost at $T_{BKT}$. This picture is further supported theoretically  by the robust correlation between these channels and the structure of the Wolff clusters, objects known to percolate at the phase coherence transition of clean systems. Moreover, our tests at $\lambda>5$ show that the two peaks (dashed lines in Fig.\ref{fig:dY}(c) move apart as disorder increases. This suggests the tempting possibility that at large enough $\lambda$ the two peaks separate out and can thus give the experimentalist a clear map of where coherence propagates, in addition to a direct determination of which of the local Josephson coupling are  below the median as opposed to above the median.
It is interesting to note that as temperature increases towards $T_{BKT}$, a significant fraction (up to approx. 40\%) of the $\delta\Upsilon$ values become negative, indicating an \emph{increase} in coherence when these bonds are cut. Such behavior is certainly possible in coherent systems, but is a departure from known results on random resistor network \cite{Arcangelis1985,Roux1991}, which is relevant to the $XY$-model at low temperatures. Again, this observation is consistent with the percolation pictures: cutting links that do not participate in the network that support the propagation of coherence, not only does not impede the global coherence, but may, in fact, increase it by giving more weight to the percolating network. The experiment suggested here and the numerical method can, in fact, be applied to any physical system involving a phase transition driven by  phase fluctuations, and may also be relevant to high-$T_C$ superconductors in part of their phase diagram. \cite{Kivelson1995}.

\textbf{Acknowledgement:} This work is supported by the ISF. Amir Erez is supported by the Adams Fellowship Program of the Israel Academy of Sciences and Humanities. Most calculations were done on the BGU HPC cluster.

\bibliography{CuttingLinks}

\begin{thebibliography}{40}%
\makeatletter
\providecommand \@ifxundefined [1]{%
 \@ifx{#1\undefined}
}%
\providecommand \@ifnum [1]{%
 \ifnum #1\expandafter \@firstoftwo
 \else \expandafter \@secondoftwo
 \fi
}%
\providecommand \@ifx [1]{%
 \ifx #1\expandafter \@firstoftwo
 \else \expandafter \@secondoftwo
 \fi
}%
\providecommand \natexlab [1]{#1}%
\providecommand \enquote  [1]{``#1''}%
\providecommand \bibnamefont  [1]{#1}%
\providecommand \bibfnamefont [1]{#1}%
\providecommand \citenamefont [1]{#1}%
\providecommand \href@noop [0]{\@secondoftwo}%
\providecommand \href [0]{\begingroup \@sanitize@url \@href}%
\providecommand \@href[1]{\@@startlink{#1}\@@href}%
\providecommand \@@href[1]{\endgroup#1\@@endlink}%
\providecommand \@sanitize@url [0]{\catcode `\\12\catcode `\$12\catcode
  `\&12\catcode `\#12\catcode `\^12\catcode `\_12\catcode `\%12\relax}%
\providecommand \@@startlink[1]{}%
\providecommand \@@endlink[0]{}%
\providecommand \url  [0]{\begingroup\@sanitize@url \@url }%
\providecommand \@url [1]{\endgroup\@href {#1}{\urlprefix }}%
\providecommand \urlprefix  [0]{URL }%
\providecommand \Eprint [0]{\href }%
\providecommand \doibase [0]{http://dx.doi.org/}%
\providecommand \selectlanguage [0]{\@gobble}%
\providecommand \bibinfo  [0]{\@secondoftwo}%
\providecommand \bibfield  [0]{\@secondoftwo}%
\providecommand \translation [1]{[#1]}%
\providecommand \BibitemOpen [0]{}%
\providecommand \bibitemStop [0]{}%
\providecommand \bibitemNoStop [0]{.\EOS\space}%
\providecommand \EOS [0]{\spacefactor3000\relax}%
\providecommand \BibitemShut  [1]{\csname bibitem#1\endcsname}%
\let\auto@bib@innerbib\@empty
\bibitem [{\citenamefont {Mermin}\ and\ \citenamefont
  {Wagner}(1966)}]{Mermin1966}%
  \BibitemOpen
  \bibfield  {author} {\bibinfo {author} {\bibfnamefont {N.~D.}\ \bibnamefont
  {Mermin}}\ and\ \bibinfo {author} {\bibfnamefont {H.}~\bibnamefont
  {Wagner}},\ }\href {http://link.aps.org/doi/10.1103/PhysRevLett.17.1133}
  {\bibfield  {journal} {\bibinfo  {journal} {Phys. Rev. Lett.}\ }\textbf
  {\bibinfo {volume} {17}},\ \bibinfo {pages} {1133} (\bibinfo {year}
  {1966})}\BibitemShut {NoStop}%
\bibitem [{\citenamefont {Berezinskiǐ}(1972)}]{Berezinskii1972}%
  \BibitemOpen
  \bibfield  {author} {\bibinfo {author} {\bibfnamefont {V.~L.}\ \bibnamefont
  {Berezinskiǐ}},\ }\href {http://adsabs.harvard.edu/abs/1972JETP...34..610B}
  {\bibfield  {journal} {\bibinfo  {journal} {Soviet Journal of Experimental
  and Theoretical Physics}\ }\textbf {\bibinfo {volume} {34}},\ \bibinfo
  {pages} {610} (\bibinfo {year} {1972})}\BibitemShut {NoStop}%
\bibitem [{\citenamefont {Kosterlitz}\ and\ \citenamefont
  {Thouless}(1973)}]{Kosterlitz1973}%
  \BibitemOpen
  \bibfield  {author} {\bibinfo {author} {\bibfnamefont {J.~M.}\ \bibnamefont
  {Kosterlitz}}\ and\ \bibinfo {author} {\bibfnamefont {D.~J.}\ \bibnamefont
  {Thouless}},\ }\href {http://stacks.iop.org/0022-3719/6/i=7/a=010} {\bibfield
   {journal} {\bibinfo  {journal} {Journal of Physics C: Solid State Physics}\
  }\textbf {\bibinfo {volume} {6}},\ \bibinfo {pages} {1181} (\bibinfo {year}
  {1973})}\BibitemShut {NoStop}%
\bibitem [{\citenamefont {Kosterlitz}(1974)}]{Kosterlitz1974}%
  \BibitemOpen
  \bibfield  {author} {\bibinfo {author} {\bibfnamefont {J.~M.}\ \bibnamefont
  {Kosterlitz}},\ }\href {http://stacks.iop.org/0022-3719/7/i=6/a=005}
  {\bibfield  {journal} {\bibinfo  {journal} {Journal of Physics C: Solid State
  Physics}\ }\textbf {\bibinfo {volume} {7}},\ \bibinfo {pages} {1046}
  (\bibinfo {year} {1974})}\BibitemShut {NoStop}%
\bibitem [{\citenamefont {Beasley}\ \emph {et~al.}(1979)\citenamefont
  {Beasley}, \citenamefont {Mooij},\ and\ \citenamefont
  {Orlando}}]{Beasley1979}%
  \BibitemOpen
  \bibfield  {author} {\bibinfo {author} {\bibfnamefont {M.~R.}\ \bibnamefont
  {Beasley}}, \bibinfo {author} {\bibfnamefont {J.~E.}\ \bibnamefont {Mooij}},
  \ and\ \bibinfo {author} {\bibfnamefont {T.~P.}\ \bibnamefont {Orlando}},\
  }\href {http://link.aps.org/doi/10.1103/PhysRevLett.42.1165} {\bibfield
  {journal} {\bibinfo  {journal} {Phys. Rev. Lett.}\ }\textbf {\bibinfo
  {volume} {42}},\ \bibinfo {pages} {1165} (\bibinfo {year}
  {1979})}\BibitemShut {NoStop}%
\bibitem [{\citenamefont {Fisher}\ \emph {et~al.}(1973)\citenamefont {Fisher},
  \citenamefont {Barber},\ and\ \citenamefont {Jasnow}}]{Fisher1973}%
  \BibitemOpen
  \bibfield  {author} {\bibinfo {author} {\bibfnamefont {M.~E.}\ \bibnamefont
  {Fisher}}, \bibinfo {author} {\bibfnamefont {M.~N.}\ \bibnamefont {Barber}},
  \ and\ \bibinfo {author} {\bibfnamefont {D.}~\bibnamefont {Jasnow}},\ }\href
  {http://link.aps.org/doi/10.1103/PhysRevA.8.1111} {\bibfield  {journal}
  {\bibinfo  {journal} {Phys. Rev. A}\ }\textbf {\bibinfo {volume} {8}},\
  \bibinfo {pages} {1111} (\bibinfo {year} {1973})}\BibitemShut {NoStop}%
\bibitem [{\citenamefont {Goldman}(2010)}]{Goldman2010}%
  \BibitemOpen
  \bibfield  {author} {\bibinfo {author} {\bibfnamefont {A.~M.}\ \bibnamefont
  {Goldman}},\ }\bibfield  {booktitle} {\emph {\bibinfo {booktitle}
  {International Journal of Modern Physics B}},\ }\href
  {http://dx.doi.org/10.1142/S0217979210056451} {\bibfield  {journal} {\bibinfo
   {journal} {Int. J. Mod. Phys. B}\ }\textbf {\bibinfo {volume} {24}},\
  \bibinfo {pages} {4081} (\bibinfo {year} {2010})}\BibitemShut {NoStop}%
\bibitem [{\citenamefont {Spivak}\ and\ \citenamefont
  {Zhou}(1995)}]{Spivak1995}%
  \BibitemOpen
  \bibfield  {author} {\bibinfo {author} {\bibfnamefont {B.}~\bibnamefont
  {Spivak}}\ and\ \bibinfo {author} {\bibfnamefont {F.}~\bibnamefont {Zhou}},\
  }\href {http://link.aps.org/doi/10.1103/PhysRevLett.74.2800} {\bibfield
  {journal} {\bibinfo  {journal} {Phys. Rev. Lett.}\ }\textbf {\bibinfo
  {volume} {74}},\ \bibinfo {pages} {2800} (\bibinfo {year}
  {1995})}\BibitemShut {NoStop}%
\bibitem [{\citenamefont {Zhou}\ and\ \citenamefont {Spivak}(1998)}]{Zhou1998}%
  \BibitemOpen
  \bibfield  {author} {\bibinfo {author} {\bibfnamefont {F.}~\bibnamefont
  {Zhou}}\ and\ \bibinfo {author} {\bibfnamefont {B.}~\bibnamefont {Spivak}},\
  }\href {http://link.aps.org/doi/10.1103/PhysRevLett.80.5647} {\bibfield
  {journal} {\bibinfo  {journal} {Phys. Rev. Lett.}\ }\textbf {\bibinfo
  {volume} {80}},\ \bibinfo {pages} {5647} (\bibinfo {year}
  {1998})}\BibitemShut {NoStop}%
\bibitem [{\citenamefont {Galitski}\ and\ \citenamefont
  {Larkin}(2001)}]{Galitski2001}%
  \BibitemOpen
  \bibfield  {author} {\bibinfo {author} {\bibfnamefont {V.~M.}\ \bibnamefont
  {Galitski}}\ and\ \bibinfo {author} {\bibfnamefont {A.~I.}\ \bibnamefont
  {Larkin}},\ }\href {http://link.aps.org/doi/10.1103/PhysRevLett.87.087001}
  {\bibfield  {journal} {\bibinfo  {journal} {Phys. Rev. Lett.}\ }\textbf
  {\bibinfo {volume} {87}},\ \bibinfo {pages} {087001} (\bibinfo {year}
  {2001})}\BibitemShut {NoStop}%
\bibitem [{\citenamefont {Ghosal}\ \emph {et~al.}(2001)\citenamefont {Ghosal},
  \citenamefont {Randeria},\ and\ \citenamefont {Trivedi}}]{Ghosal2001}%
  \BibitemOpen
  \bibfield  {author} {\bibinfo {author} {\bibfnamefont {A.}~\bibnamefont
  {Ghosal}}, \bibinfo {author} {\bibfnamefont {M.}~\bibnamefont {Randeria}}, \
  and\ \bibinfo {author} {\bibfnamefont {N.}~\bibnamefont {Trivedi}},\ }\href
  {http://link.aps.org/doi/10.1103/PhysRevB.65.014501} {\bibfield  {journal}
  {\bibinfo  {journal} {Phys. Rev. B}\ }\textbf {\bibinfo {volume} {65}},\
  \bibinfo {pages} {014501} (\bibinfo {year} {2001})}\BibitemShut {NoStop}%
\bibitem [{\citenamefont {Dubi}\ \emph {et~al.}(2007)\citenamefont {Dubi},
  \citenamefont {Meir},\ and\ \citenamefont {Avishai}}]{Dubi2007}%
  \BibitemOpen
  \bibfield  {author} {\bibinfo {author} {\bibfnamefont {Y.}~\bibnamefont
  {Dubi}}, \bibinfo {author} {\bibfnamefont {Y.}~\bibnamefont {Meir}}, \ and\
  \bibinfo {author} {\bibfnamefont {Y.}~\bibnamefont {Avishai}},\ }\href
  {http://dx.doi.org/10.1038/nature06180} {\bibfield  {journal} {\bibinfo
  {journal} {Nature}\ }\textbf {\bibinfo {volume} {449}},\ \bibinfo {pages}
  {876} (\bibinfo {year} {2007})}\BibitemShut {NoStop}%
\bibitem [{\citenamefont {Erez}\ and\ \citenamefont
  {Meir}(2010)}]{ErezMeir2010}%
  \BibitemOpen
  \bibfield  {author} {\bibinfo {author} {\bibfnamefont {A.}~\bibnamefont
  {Erez}}\ and\ \bibinfo {author} {\bibfnamefont {Y.}~\bibnamefont {Meir}},\
  }\href {http://stacks.iop.org/0295-5075/91/i=4/a=47003} {\bibfield  {journal}
  {\bibinfo  {journal} {EPL (Europhysics Letters)}\ }\textbf {\bibinfo {volume}
  {91}},\ \bibinfo {pages} {47003} (\bibinfo {year} {2010})}\BibitemShut
  {NoStop}%
\bibitem [{\citenamefont {Kowal}\ and\ \citenamefont
  {Ovadyahu}(1994)}]{Kowal1994}%
  \BibitemOpen
  \bibfield  {author} {\bibinfo {author} {\bibfnamefont {D.}~\bibnamefont
  {Kowal}}\ and\ \bibinfo {author} {\bibfnamefont {Z.}~\bibnamefont
  {Ovadyahu}},\ }\href
  {http://www.sciencedirect.com/science/article/pii/0038109894902429}
  {\bibfield  {journal} {\bibinfo  {journal} {Solid State Communications}\
  }\textbf {\bibinfo {volume} {90}},\ \bibinfo {pages} {783} (\bibinfo {year}
  {1994})}\BibitemShut {NoStop}%
\bibitem [{\citenamefont {Howald}\ \emph {et~al.}(2001)\citenamefont {Howald},
  \citenamefont {Fournier},\ and\ \citenamefont {Kapitulnik}}]{Howald2001}%
  \BibitemOpen
  \bibfield  {author} {\bibinfo {author} {\bibfnamefont {C.}~\bibnamefont
  {Howald}}, \bibinfo {author} {\bibfnamefont {P.}~\bibnamefont {Fournier}}, \
  and\ \bibinfo {author} {\bibfnamefont {A.}~\bibnamefont {Kapitulnik}},\
  }\href {http://link.aps.org/doi/10.1103/PhysRevB.64.100504} {\bibfield
  {journal} {\bibinfo  {journal} {Phys. Rev. B}\ }\textbf {\bibinfo {volume}
  {64}},\ \bibinfo {pages} {100504} (\bibinfo {year} {2001})}\BibitemShut
  {NoStop}%
\bibitem [{\citenamefont {Sacepe}\ \emph {et~al.}(2008)\citenamefont {Sacepe},
  \citenamefont {Chapelier}, \citenamefont {Baturina}, \citenamefont {Vinokur},
  \citenamefont {Baklanov},\ and\ \citenamefont {Sanquer}}]{Sacepe2008}%
  \BibitemOpen
  \bibfield  {author} {\bibinfo {author} {\bibfnamefont {B.}~\bibnamefont
  {Sacepe}}, \bibinfo {author} {\bibfnamefont {C.}~\bibnamefont {Chapelier}},
  \bibinfo {author} {\bibfnamefont {T.~I.}\ \bibnamefont {Baturina}}, \bibinfo
  {author} {\bibfnamefont {V.~M.}\ \bibnamefont {Vinokur}}, \bibinfo {author}
  {\bibfnamefont {M.~R.}\ \bibnamefont {Baklanov}}, \ and\ \bibinfo {author}
  {\bibfnamefont {M.}~\bibnamefont {Sanquer}},\ }\href
  {http://link.aps.org/doi/10.1103/PhysRevLett.101.157006} {\bibfield
  {journal} {\bibinfo  {journal} {Phys. Rev. Lett.}\ }\textbf {\bibinfo
  {volume} {101}},\ \bibinfo {pages} {157006} (\bibinfo {year}
  {2008})}\BibitemShut {NoStop}%
\bibitem [{\citenamefont {Sacepe}\ \emph {et~al.}(2011)\citenamefont {Sacepe},
  \citenamefont {Dubouchet}, \citenamefont {Chapelier}, \citenamefont
  {Sanquer}, \citenamefont {Ovadia}, \citenamefont {Shahar}, \citenamefont
  {Feigelman},\ and\ \citenamefont {Ioffe}}]{Sacepe2011}%
  \BibitemOpen
  \bibfield  {author} {\bibinfo {author} {\bibfnamefont {B.}~\bibnamefont
  {Sacepe}}, \bibinfo {author} {\bibfnamefont {T.}~\bibnamefont {Dubouchet}},
  \bibinfo {author} {\bibfnamefont {C.}~\bibnamefont {Chapelier}}, \bibinfo
  {author} {\bibfnamefont {M.}~\bibnamefont {Sanquer}}, \bibinfo {author}
  {\bibfnamefont {M.}~\bibnamefont {Ovadia}}, \bibinfo {author} {\bibfnamefont
  {D.}~\bibnamefont {Shahar}}, \bibinfo {author} {\bibfnamefont
  {M.}~\bibnamefont {Feigelman}}, \ and\ \bibinfo {author} {\bibfnamefont
  {L.}~\bibnamefont {Ioffe}},\ }\href {http://dx.doi.org/10.1038/nphys1892}
  {\bibfield  {journal} {\bibinfo  {journal} {Nat Phys}\ }\textbf {\bibinfo
  {volume} {7}},\ \bibinfo {pages} {239} (\bibinfo {year} {2011})}\BibitemShut
  {NoStop}%
\bibitem [{\citenamefont {Scalettar}\ \emph {et~al.}(1999)\citenamefont
  {Scalettar}, \citenamefont {Trivedi},\ and\ \citenamefont
  {Huscroft}}]{Scalettar1999}%
  \BibitemOpen
  \bibfield  {author} {\bibinfo {author} {\bibfnamefont {R.~T.}\ \bibnamefont
  {Scalettar}}, \bibinfo {author} {\bibfnamefont {N.}~\bibnamefont {Trivedi}},
  \ and\ \bibinfo {author} {\bibfnamefont {C.}~\bibnamefont {Huscroft}},\
  }\href {http://link.aps.org/doi/10.1103/PhysRevB.59.4364} {\bibfield
  {journal} {\bibinfo  {journal} {Phys. Rev. B}\ }\textbf {\bibinfo {volume}
  {59}},\ \bibinfo {pages} {4364} (\bibinfo {year} {1999})}\BibitemShut
  {NoStop}%
\bibitem [{\citenamefont {Yazdani}\ and\ \citenamefont
  {Kapitulnik}(1995)}]{Yazdani1995}%
  \BibitemOpen
  \bibfield  {author} {\bibinfo {author} {\bibfnamefont {A.}~\bibnamefont
  {Yazdani}}\ and\ \bibinfo {author} {\bibfnamefont {A.}~\bibnamefont
  {Kapitulnik}},\ }\href {http://link.aps.org/doi/10.1103/PhysRevLett.74.3037}
  {\bibfield  {journal} {\bibinfo  {journal} {Phys. Rev. Lett.}\ }\textbf
  {\bibinfo {volume} {74}},\ \bibinfo {pages} {3037} (\bibinfo {year}
  {1995})}\BibitemShut {NoStop}%
\bibitem [{\citenamefont {Gantmakher}\ \emph {et~al.}(1998)\citenamefont
  {Gantmakher}, \citenamefont {Golubkov}, \citenamefont {Dolgopolov},
  \citenamefont {Tsydynzhapov},\ and\ \citenamefont
  {Shashkin}}]{Gantmakher1998}%
  \BibitemOpen
  \bibfield  {author} {\bibinfo {author} {\bibfnamefont {V.}~\bibnamefont
  {Gantmakher}}, \bibinfo {author} {\bibfnamefont {M.}~\bibnamefont
  {Golubkov}}, \bibinfo {author} {\bibfnamefont {V.}~\bibnamefont
  {Dolgopolov}}, \bibinfo {author} {\bibfnamefont {G.}~\bibnamefont
  {Tsydynzhapov}}, \ and\ \bibinfo {author} {\bibfnamefont {A.}~\bibnamefont
  {Shashkin}},\ }\bibfield  {booktitle} {\emph {\bibinfo {booktitle} {Journal
  of Experimental and Theoretical Physics Letters}},\ }\href
  {http://dx.doi.org/10.1134/1.567874} {\ \textbf {\bibinfo {volume} {68}},\
  \bibinfo {pages} {363} (\bibinfo {year} {1998})}\BibitemShut {NoStop}%
\bibitem [{\citenamefont {Crane}\ \emph {et~al.}(2007)\citenamefont {Crane},
  \citenamefont {Armitage}, \citenamefont {Johansson}, \citenamefont
  {Sambandamurthy}, \citenamefont {Shahar},\ and\ \citenamefont
  {Gruner}}]{Crane2007}%
  \BibitemOpen
  \bibfield  {author} {\bibinfo {author} {\bibfnamefont {R.}~\bibnamefont
  {Crane}}, \bibinfo {author} {\bibfnamefont {N.~P.}\ \bibnamefont {Armitage}},
  \bibinfo {author} {\bibfnamefont {A.}~\bibnamefont {Johansson}}, \bibinfo
  {author} {\bibfnamefont {G.}~\bibnamefont {Sambandamurthy}}, \bibinfo
  {author} {\bibfnamefont {D.}~\bibnamefont {Shahar}}, \ and\ \bibinfo {author}
  {\bibfnamefont {G.}~\bibnamefont {Gruner}},\ }\href
  {http://link.aps.org/doi/10.1103/PhysRevB.75.184530} {\bibfield  {journal}
  {\bibinfo  {journal} {Phys. Rev. B}\ }\textbf {\bibinfo {volume} {75}},\
  \bibinfo {pages} {184530} (\bibinfo {year} {2007})}\BibitemShut {NoStop}%
\bibitem [{\citenamefont {Stewart}\ \emph {et~al.}(2007)\citenamefont
  {Stewart}, \citenamefont {Yin}, \citenamefont {Xu},\ and\ \citenamefont
  {Valles}}]{Stewart2007}%
  \BibitemOpen
  \bibfield  {author} {\bibinfo {author} {\bibfnamefont {M.~D.}\ \bibnamefont
  {Stewart}}, \bibinfo {author} {\bibfnamefont {A.}~\bibnamefont {Yin}},
  \bibinfo {author} {\bibfnamefont {J.~M.}\ \bibnamefont {Xu}}, \ and\ \bibinfo
  {author} {\bibfnamefont {J.~M.}\ \bibnamefont {Valles}},\ }\href {\doibase
  10.1126/science.1149587} {\bibfield  {journal} {\bibinfo  {journal}
  {Science}\ }\textbf {\bibinfo {volume} {318}},\ \bibinfo {pages} {1273}
  (\bibinfo {year} {2007})}\BibitemShut {NoStop}%
\bibitem [{\citenamefont {Stewart}\ \emph {et~al.}(2008)\citenamefont
  {Stewart}, \citenamefont {Yin}, \citenamefont {Xu},\ and\ \citenamefont
  {Valles}}]{Stewart2008}%
  \BibitemOpen
  \bibfield  {author} {\bibinfo {author} {\bibfnamefont {J.}~\bibnamefont
  {Stewart}, \bibfnamefont {M.~D.}}, \bibinfo {author} {\bibfnamefont
  {A.}~\bibnamefont {Yin}}, \bibinfo {author} {\bibfnamefont {J.~M.}\
  \bibnamefont {Xu}}, \ and\ \bibinfo {author} {\bibfnamefont {J.}~\bibnamefont
  {Valles}, \bibfnamefont {J.~M.}},\ }\href
  {http://link.aps.org/doi/10.1103/PhysRevB.77.140501} {\bibfield  {journal}
  {\bibinfo  {journal} {Phys. Rev. B}\ }\textbf {\bibinfo {volume} {77}},\
  \bibinfo {pages} {140501} (\bibinfo {year} {2008})}\BibitemShut {NoStop}%
\bibitem [{\citenamefont {Nguyen}\ \emph {et~al.}(2009)\citenamefont {Nguyen},
  \citenamefont {Hollen}, \citenamefont {Stewart}, \citenamefont {Shainline},
  \citenamefont {Yin}, \citenamefont {Xu},\ and\ \citenamefont
  {Valles}}]{Nguyen2009}%
  \BibitemOpen
  \bibfield  {author} {\bibinfo {author} {\bibfnamefont {H.~Q.}\ \bibnamefont
  {Nguyen}}, \bibinfo {author} {\bibfnamefont {S.~M.}\ \bibnamefont {Hollen}},
  \bibinfo {author} {\bibfnamefont {J.}~\bibnamefont {Stewart}, \bibfnamefont
  {M.~D.}}, \bibinfo {author} {\bibfnamefont {J.}~\bibnamefont {Shainline}},
  \bibinfo {author} {\bibfnamefont {A.}~\bibnamefont {Yin}}, \bibinfo {author}
  {\bibfnamefont {J.~M.}\ \bibnamefont {Xu}}, \ and\ \bibinfo {author}
  {\bibfnamefont {J.}~\bibnamefont {Valles}, \bibfnamefont {J.~M.}},\ }\href
  {http://link.aps.org/doi/10.1103/PhysRevLett.103.157001} {\bibfield
  {journal} {\bibinfo  {journal} {Phys. Rev. Lett.}\ }\textbf {\bibinfo
  {volume} {103}},\ \bibinfo {pages} {157001} (\bibinfo {year}
  {2009})}\BibitemShut {NoStop}%
\bibitem [{\citenamefont {Harris}(1974)}]{Harris1974}%
  \BibitemOpen
  \bibfield  {author} {\bibinfo {author} {\bibfnamefont {A.~B.}\ \bibnamefont
  {Harris}},\ }\href {http://stacks.iop.org/0022-3719/7/i=9/a=009} {\enquote
  {\bibinfo {title} {Effect of random defects on the critical behaviour of
  ising models},}\ } (\bibinfo {year} {1974})\BibitemShut {NoStop}%
\bibitem [{\citenamefont {Misra}\ \emph {et~al.}(2013)\citenamefont {Misra},
  \citenamefont {Urban}, \citenamefont {Kim}, \citenamefont {Sambandamurthy},\
  and\ \citenamefont {Yazdani}}]{Misra2013}%
  \BibitemOpen
  \bibfield  {author} {\bibinfo {author} {\bibfnamefont {S.}~\bibnamefont
  {Misra}}, \bibinfo {author} {\bibfnamefont {L.}~\bibnamefont {Urban}},
  \bibinfo {author} {\bibfnamefont {M.}~\bibnamefont {Kim}}, \bibinfo {author}
  {\bibfnamefont {G.}~\bibnamefont {Sambandamurthy}}, \ and\ \bibinfo {author}
  {\bibfnamefont {A.}~\bibnamefont {Yazdani}},\ }\href
  {http://link.aps.org/doi/10.1103/PhysRevLett.110.037002} {\bibfield
  {journal} {\bibinfo  {journal} {Phys. Rev. Lett.}\ }\textbf {\bibinfo
  {volume} {110}},\ \bibinfo {pages} {037002} (\bibinfo {year}
  {2013})}\BibitemShut {NoStop}%
\bibitem [{\citenamefont {Benfatto}\ \emph {et~al.}(2007)\citenamefont
  {Benfatto}, \citenamefont {Castellani},\ and\ \citenamefont
  {Giamarchi}}]{Benfatto2007}%
  \BibitemOpen
  \bibfield  {author} {\bibinfo {author} {\bibfnamefont {L.}~\bibnamefont
  {Benfatto}}, \bibinfo {author} {\bibfnamefont {C.}~\bibnamefont
  {Castellani}}, \ and\ \bibinfo {author} {\bibfnamefont {T.}~\bibnamefont
  {Giamarchi}},\ }\href {http://link.aps.org/doi/10.1103/PhysRevLett.98.117008}
  {\bibfield  {journal} {\bibinfo  {journal} {Phys. Rev. Lett.}\ }\textbf
  {\bibinfo {volume} {98}},\ \bibinfo {pages} {117008} (\bibinfo {year}
  {2007})}\BibitemShut {NoStop}%
\bibitem [{\citenamefont {Topinka}\ \emph {et~al.}(2001)\citenamefont
  {Topinka}, \citenamefont {LeRoy}, \citenamefont {Westervelt}, \citenamefont
  {Shaw}, \citenamefont {Fleischmann}, \citenamefont {Heller}, \citenamefont
  {Maranowski},\ and\ \citenamefont {Gossard}}]{Topinka2001}%
  \BibitemOpen
  \bibfield  {author} {\bibinfo {author} {\bibfnamefont {M.~A.}\ \bibnamefont
  {Topinka}}, \bibinfo {author} {\bibfnamefont {B.~J.}\ \bibnamefont {LeRoy}},
  \bibinfo {author} {\bibfnamefont {R.~M.}\ \bibnamefont {Westervelt}},
  \bibinfo {author} {\bibfnamefont {S.~E.~J.}\ \bibnamefont {Shaw}}, \bibinfo
  {author} {\bibfnamefont {R.}~\bibnamefont {Fleischmann}}, \bibinfo {author}
  {\bibfnamefont {E.~J.}\ \bibnamefont {Heller}}, \bibinfo {author}
  {\bibfnamefont {K.~D.}\ \bibnamefont {Maranowski}}, \ and\ \bibinfo {author}
  {\bibfnamefont {A.~C.}\ \bibnamefont {Gossard}},\ }\href
  {http://dx.doi.org/10.1038/35065553} {\bibfield  {journal} {\bibinfo
  {journal} {Nature}\ }\textbf {\bibinfo {volume} {410}},\ \bibinfo {pages}
  {183} (\bibinfo {year} {2001})}\BibitemShut {NoStop}%
\bibitem [{\citenamefont {Kiccaronin}\ \emph {et~al.}(2004)\citenamefont
  {Kiccaronin}, \citenamefont {Pioda}, \citenamefont {Ihn}, \citenamefont
  {Ensslin}, \citenamefont {Driscoll},\ and\ \citenamefont
  {Gossard}}]{Kiccaronin2004}%
  \BibitemOpen
  \bibfield  {author} {\bibinfo {author} {\bibfnamefont {S.}~\bibnamefont
  {Kiccaronin}}, \bibinfo {author} {\bibfnamefont {A.}~\bibnamefont {Pioda}},
  \bibinfo {author} {\bibfnamefont {T.}~\bibnamefont {Ihn}}, \bibinfo {author}
  {\bibfnamefont {K.}~\bibnamefont {Ensslin}}, \bibinfo {author} {\bibfnamefont
  {D.~C.}\ \bibnamefont {Driscoll}}, \ and\ \bibinfo {author} {\bibfnamefont
  {A.~C.}\ \bibnamefont {Gossard}},\ }\href
  {http://link.aps.org/doi/10.1103/PhysRevB.70.205302} {\bibfield  {journal}
  {\bibinfo  {journal} {Phys. Rev. B}\ }\textbf {\bibinfo {volume} {70}},\
  \bibinfo {pages} {205302} (\bibinfo {year} {2004})}\BibitemShut {NoStop}%
\bibitem [{\citenamefont {Jura}\ \emph {et~al.}(2007)\citenamefont {Jura},
  \citenamefont {Topinka}, \citenamefont {Urban}, \citenamefont {Yazdani},
  \citenamefont {Shtrikman}, \citenamefont {Pfeiffer}, \citenamefont {West},\
  and\ \citenamefont {Goldhaber-Gordon}}]{Jura2007}%
  \BibitemOpen
  \bibfield  {author} {\bibinfo {author} {\bibfnamefont {M.~P.}\ \bibnamefont
  {Jura}}, \bibinfo {author} {\bibfnamefont {M.~A.}\ \bibnamefont {Topinka}},
  \bibinfo {author} {\bibfnamefont {L.}~\bibnamefont {Urban}}, \bibinfo
  {author} {\bibfnamefont {A.}~\bibnamefont {Yazdani}}, \bibinfo {author}
  {\bibfnamefont {H.}~\bibnamefont {Shtrikman}}, \bibinfo {author}
  {\bibfnamefont {L.~N.}\ \bibnamefont {Pfeiffer}}, \bibinfo {author}
  {\bibfnamefont {K.~W.}\ \bibnamefont {West}}, \ and\ \bibinfo {author}
  {\bibfnamefont {D.}~\bibnamefont {Goldhaber-Gordon}},\ }\href
  {http://adsabs.harvard.edu/abs/2007NatPh...3..841J} {\bibfield  {journal}
  {\bibinfo  {journal} {Nature Physics}\ }\textbf {\bibinfo {volume} {3}},\
  \bibinfo {pages} {841} (\bibinfo {year} {2007})}\BibitemShut {NoStop}%
\bibitem [{\citenamefont {Wolff}(1989)}]{Wolff1989}%
  \BibitemOpen
  \bibfield  {author} {\bibinfo {author} {\bibfnamefont {U.}~\bibnamefont
  {Wolff}},\ }\href {http://link.aps.org/doi/10.1103/PhysRevLett.62.361}
  {\bibfield  {journal} {\bibinfo  {journal} {Phys. Rev. Lett.}\ }\textbf
  {\bibinfo {volume} {62}},\ \bibinfo {pages} {361} (\bibinfo {year}
  {1989})}\BibitemShut {NoStop}%
\bibitem [{\citenamefont {Chayes}(1998)}]{Chayes1998}%
  \BibitemOpen
  \bibfield  {author} {\bibinfo {author} {\bibfnamefont {L.}~\bibnamefont
  {Chayes}},\ }\href {http://dx.doi.org/10.1007/s002200050466} {\bibfield
  {journal} {\bibinfo  {journal} {Communications in Mathematical Physics}\
  }\textbf {\bibinfo {volume} {197}},\ \bibinfo {pages} {623} (\bibinfo {year}
  {1998})},\ \bibinfo {note} {10.1007/s002200050466}\BibitemShut {NoStop}%
\bibitem [{\citenamefont {Emery}\ and\ \citenamefont
  {Kivelson}(1995)}]{Kivelson1995}%
  \BibitemOpen
  \bibfield  {author} {\bibinfo {author} {\bibfnamefont {V.~J.}\ \bibnamefont
  {Emery}}\ and\ \bibinfo {author} {\bibfnamefont {S.~A.}\ \bibnamefont
  {Kivelson}},\ }\href {\doibase 10.1038/374434a0} {\bibfield  {journal}
  {\bibinfo  {journal} {Nature}\ }\textbf {\bibinfo {volume} {374}},\ \bibinfo
  {pages} {434} (\bibinfo {year} {1995})}\BibitemShut {NoStop}%
\bibitem [{\citenamefont {Carlson}\ \emph {et~al.}(2008)\citenamefont
  {Carlson}, \citenamefont {Emery}, \citenamefont {Kivelson},\ and\
  \citenamefont {Orgad}}]{Carlson2008}%
  \BibitemOpen
  \bibfield  {author} {\bibinfo {author} {\bibfnamefont {E.}~\bibnamefont
  {Carlson}}, \bibinfo {author} {\bibfnamefont {V.}~\bibnamefont {Emery}},
  \bibinfo {author} {\bibfnamefont {S.}~\bibnamefont {Kivelson}}, \ and\
  \bibinfo {author} {\bibfnamefont {D.}~\bibnamefont {Orgad}},\ }in\ \href
  {http://dx.doi.org/10.1007/978-3-540-73253-2_21} {\emph {\bibinfo {booktitle}
  {Superconductivity}}},\ \bibinfo {editor} {edited by\ \bibinfo {editor}
  {\bibfnamefont {K.}~\bibnamefont {Bennemann}}\ and\ \bibinfo {editor}
  {\bibfnamefont {J.}~\bibnamefont {Ketterson}}}\ (\bibinfo  {publisher}
  {Springer Berlin Heidelberg},\ \bibinfo {year} {2008})\ pp.\ \bibinfo {pages}
  {1225--1348--}\BibitemShut {NoStop}%
\bibitem [{\citenamefont {Ambegaokar}\ \emph {et~al.}(1971)\citenamefont
  {Ambegaokar}, \citenamefont {Halperin},\ and\ \citenamefont
  {Langer}}]{Ambegaokar1971}%
  \BibitemOpen
  \bibfield  {author} {\bibinfo {author} {\bibfnamefont {V.}~\bibnamefont
  {Ambegaokar}}, \bibinfo {author} {\bibfnamefont {B.~I.}\ \bibnamefont
  {Halperin}}, \ and\ \bibinfo {author} {\bibfnamefont {J.~S.}\ \bibnamefont
  {Langer}},\ }\href {http://link.aps.org/doi/10.1103/PhysRevB.4.2612}
  {\bibfield  {journal} {\bibinfo  {journal} {Phys. Rev. B}\ }\textbf {\bibinfo
  {volume} {4}},\ \bibinfo {pages} {2612} (\bibinfo {year} {1971})}\BibitemShut
  {NoStop}%
\bibitem [{\citenamefont {Doussal}(1989)}]{Doussal1989}%
  \BibitemOpen
  \bibfield  {author} {\bibinfo {author} {\bibfnamefont {P.~L.}\ \bibnamefont
  {Doussal}},\ }\href {http://link.aps.org/doi/10.1103/PhysRevB.39.881}
  {\bibfield  {journal} {\bibinfo  {journal} {Phys. Rev. B}\ }\textbf {\bibinfo
  {volume} {39}},\ \bibinfo {pages} {881} (\bibinfo {year} {1989})}\BibitemShut
  {NoStop}%
\bibitem [{\citenamefont {Fisher}(1961)}]{Fisher1961}%
  \BibitemOpen
  \bibfield  {author} {\bibinfo {author} {\bibfnamefont {M.~E.}\ \bibnamefont
  {Fisher}},\ }\href {http://dx.doi.org/10.1063/1.1703746} {\bibfield
  {journal} {\bibinfo  {journal} {J. Math. Phys.}\ }\textbf {\bibinfo {volume}
  {2}},\ \bibinfo {pages} {620} (\bibinfo {year} {1961})}\BibitemShut {NoStop}%
\bibitem [{\citenamefont {Leung}\ and\ \citenamefont
  {Henley}(1991)}]{Henley1991}%
  \BibitemOpen
  \bibfield  {author} {\bibinfo {author} {\bibfnamefont {P.~W.}\ \bibnamefont
  {Leung}}\ and\ \bibinfo {author} {\bibfnamefont {C.~L.}\ \bibnamefont
  {Henley}},\ }\href {http://link.aps.org/doi/10.1103/PhysRevB.43.752}
  {\bibfield  {journal} {\bibinfo  {journal} {Phys. Rev. B}\ }\textbf {\bibinfo
  {volume} {43}},\ \bibinfo {pages} {752} (\bibinfo {year} {1991})}\BibitemShut
  {NoStop}%
\bibitem [{\citenamefont {Arcangelis}\ \emph {et~al.}(1985)\citenamefont
  {Arcangelis}, \citenamefont {Redner},\ and\ \citenamefont
  {Coniglio}}]{Arcangelis1985}%
  \BibitemOpen
  \bibfield  {author} {\bibinfo {author} {\bibfnamefont {L.~d.}\ \bibnamefont
  {Arcangelis}}, \bibinfo {author} {\bibfnamefont {S.}~\bibnamefont {Redner}},
  \ and\ \bibinfo {author} {\bibfnamefont {A.}~\bibnamefont {Coniglio}},\
  }\href {http://link.aps.org/doi/10.1103/PhysRevB.31.4725} {\bibfield
  {journal} {\bibinfo  {journal} {Phys. Rev. B}\ }\textbf {\bibinfo {volume}
  {31}},\ \bibinfo {pages} {4725} (\bibinfo {year} {1985})}\BibitemShut
  {NoStop}%
\bibitem [{\citenamefont {Roux}\ \emph {et~al.}(1991)\citenamefont {Roux},
  \citenamefont {Rigord}, \citenamefont {Hansen},\ and\ \citenamefont
  {Hinrichsen}}]{Roux1991}%
  \BibitemOpen
  \bibfield  {author} {\bibinfo {author} {\bibfnamefont {S.}~\bibnamefont
  {Roux}}, \bibinfo {author} {\bibfnamefont {P.}~\bibnamefont {Rigord}},
  \bibinfo {author} {\bibfnamefont {A.}~\bibnamefont {Hansen}}, \ and\ \bibinfo
  {author} {\bibfnamefont {E.~L.}\ \bibnamefont {Hinrichsen}},\ }\href
  {http://link.aps.org/doi/10.1103/PhysRevB.43.10984} {\bibfield  {journal}
  {\bibinfo  {journal} {Phys. Rev. B}\ }\textbf {\bibinfo {volume} {43}},\
  \bibinfo {pages} {10984} (\bibinfo {year} {1991})}\BibitemShut {NoStop}%
\end{thebibliography}%

\end{document}